\documentclass[aps,pra,reprint]{revtex4-1}

\usepackage{amsmath,amsfonts,amssymb,amscd,amsthm,xspace}
\usepackage{graphicx}
\usepackage{epstopdf}
\usepackage{enumerate}
\usepackage{wrapfig}
\usepackage{color}
\usepackage{longtable}
\usepackage{float}
\usepackage{hyperref}

\usepackage{times}

\begin{document}


\title{Stability of self-gravitating Bose-Einstein-Condensates} 



\author{Kris Schroven}
\email[]{schroven@zarm.uni-bremen.de}
\author{Meike List}
\email[]{meike.list@zarm.uni-bremen.de}
\author{Claus L\"ammerzahl}
\email[]{claus.laemmerzahl@zarm.uni-bremen.de}
\affiliation{ZARM, University of Bremen, Am Fallturm, 28359 Bremen, Germany}


\date{\today}

\begin{abstract}
  We study the ground state and the first three radially excited states of a self-gravitating Bose-Einstein-Condensate with respect to the influence of two external parameters, the total mass and the 
  strength of interactions between particles. For this we use the so-called Gross-Pitaevskii-Newton system.
  In this context we especially determine
  the case of very high total masses where the ground state solutions of the Gross-Pitaevskii-Newton system can be approximated with the Thomas-Fermi limit. 
  Furthermore, stability properties of the computed radially excited states are examined by applying arguments of the catastrophe theory. 
\end{abstract}

\pacs{}

\maketitle 
 \section{Introduction}
  The first Bose-Einstein-Condensate (BEC) was produced in 1995 by the group of E. Cornell and C. Wiemann \cite{firstbec}, 70 years after its prediction by S. Bose and A. Einstein 1925 \cite{einstein}. Being a 
  realization of a macroscopic quantum object, BECs have several interesting properties and therefore have been extensively studied experimentally as well as theoretically.

  BECs are a dilute quantum gas with short range dipole interactions between the atoms. Thus, for ultracold temperatures they are described by means of the Gross-Pitaevskii equation (GP equation) \cite{gross,pit} 
  \begin{equation}
  i \hbar \frac{\partial}{\partial t} \psi = - \frac{\hbar^2}{2m} \Delta \psi + V \psi + g |\psi|^2 \psi 
  \end{equation}
  coupled to an external potential $V$ describing in particular the traps (e.g. a harmonic potential) \cite{pethick}. Here $\psi$ represents the wave functions of the condensate. The parameter $g$ describes 
  the self-interaction. Despite their diluteness, it is interesting to discuss self-gravitating BECs from a conceptual point of view and also from an experimental and astrophysical perspective. 

  Self-gravitating quantum systems have been proposed by R. Penrose in his discussion of quantum state reduction by gravity \cite{grq}. He considered a self-gravitating Schr\"odinger field described by the 
  Schr\"odinger-Newton (SN) equations 
  \begin{align}\begin{split}
  i \hbar \frac{\partial}{\partial t} \psi & = - \frac{\hbar^2}{2m} \Delta \psi + V\psi + m \Phi \psi \\
  \Delta \Phi & = 4 \pi G |\psi|^2 \, , \end{split}
  \label{sneq}
  \end{align}
  where $G$ is the Newton constant. This setting would dramatically change the concept of quantum mechanics, where one only assumes interactions with other external fields.

  In an astrophysical context, R. Ruffini and S. Bonazzola were the first to discuss self-gravitating bosons which are exclusively trapped in their own gravitational potentials, 
  as a concept for boson stars \cite{ruffini}. The SN equations have been studied further extensively by R. Harrison and I.M. Moroz et al. \cite{dissnewton,snb1,*snb2,*snb3,*snb4,*snb5}. 

  Here we are going to discuss self-gravitating BECs given by the Gross-Pitaevskii-Newton (GPN) system
  \begin{align}\begin{split}
  i \hbar \frac{\partial}{\partial t} \psi & = - \frac{\hbar^2}{2m} \Delta \psi + m \Phi \psi + V \psi + g |\psi|^2 \psi\\
  \Delta \Phi & = 4 \pi G |\psi|^2 \,. \end{split}
  \end{align}
  Contrary to the SN equations, a self-gravitating BEC will still have a meaning within the standard quantum mechanical framework. This is as well the case 
  for the limit $g \rightarrow 0$, which recovers the SN equations in \eqref{sneq}. In this case however $\psi$ in \eqref{sneq} has to be interpreted as wave functions of the condensate. On the experimental 
  side, one may think of creating a self-gravitating BEC through roque waves which can be excited within BEC due to the nonlinearity of the GP equation \cite{baronioetal12,Sunetal14}. The GPN system 
  also may serve as model for nonrelativistic boson stars. 
  The SN equations are the non-relativistic limit of the Einstein-Klein-Gordon (EKG) system \cite{snb4}. Analogously, the GPN system can be obtained as nonrelativistic limit of a generalized EKG system 
  \begin{align}\begin{split}
  \square \psi + U(\psi) & = 0 \\
  R_{\mu\nu} - \frac12 g_{\mu\nu} R & = T_{\mu\nu} \end{split}
  \end{align}
  with an extra potential $U$ for the Boson field describing self-interactions due to local interactions between the Bosons. $T_{\mu\nu}$ is the energy momentum tensor of the Klein-Gordon field. Usually $U(\psi)$ 
  is given by some polynomial. For the usual Klein-Gordon equation without local self-interactions we have $U(\psi) = m \psi$. Such EKG systems have been extensively studied as model for relativistic boson stars, 
  see \cite{Jetzer:1991jr,SchunckMielke03} for reviews. 

  Giant self-gravitating BECs have been suggested as candidates for dark matter (DM) halos (e.g. \cite{baldeschi,dmx,dmkorean,reldm,dmboehmer}). While C. G. B\"ohmer and T. Harko discussed the Thomas-Fermi limit 
  (TF limit), P.-H. Chavanis discussed the full GPN system numerically \cite{cnum} and analytically \cite{cana} for ground state solutions. 

  The nature of DM is one of the major quests in cosmology and theoretical physics. Among others the flatness of observed galaxy rotation curves can not be explained by Newtonian gravity or standard general 
  relativity, if only the visible matter is considered. As a result one assumes the existence of invisible dark matter which forms a spherical halo around galaxies. There are a number of suggested dark matter 
  models, most popular the $\Lambda$ cold dark matter ($\Lambda$CDM) model which comprises weakly interacting massive particles (WIMPs). But there arise difficulties to explain the observed distribution of the 
  invisible matter at galactic centers (in the scales of the order of 1 kps and smaller) with the CDM model. It leads to cuspy density profiles \cite{cuspproblem} instead.
  In the context of the CDM model self-gravitating BECs are discussed as a dark matter candidate which solves the occurring cusp problem. The quantum properties of self-gravitating BECs lead to a repulsive force 
  due to the Heisenberg uncertainty principle which prevents the forming of cusps in the density profile \cite{harkocusp}. Rotation curves induced by self-gravitating BECs have been compared to observed galaxy 
  rotation curves using different relativistic \cite{reldm} and non-relativistic models (e.g. using the TF limit \cite{dmboehmer}, neglecting particle interactions \cite{dmkorean} or discussing rotating 
  self-gravitating BECs \cite{rotate}) and seem to approach the observed curves.

  Now the question arises, whether the obtained solutions for a self-gravitating BEC are actually stable. 
  The existence of a sort of Jeans instability for an infinite spatially homogeneous distribution of self-gravitating bosons could be shown in \cite{khlopov,bianchi} and in \cite{cana} (taking into account short-range interactions).
  Discussing self-gravitating BECs, it was shown in case of a relativistic treatment that for a potential $U(\psi)$ of the 
  form $U_1(\psi)=\lambda \left( \psi^6 -a\psi^4+b\psi^2 \right)$ or $U_2(\psi)=\lambda' \left( \psi^4 -a'\psi^3+b'\psi^2 \right)$ for $\lambda,\lambda',a,a',b,b'>0$, stable and unstable ground state solutions 
  exist \cite{bstar12,sak}. If self-interactions between particles are neglected the found ground state solutions are stable \cite{miniboson}. Radially excited states on the other hand seem to be unstable either 
  way, if the potential $U_2$ or no self-interaction is considered \cite{Balakrishna,miniboson}. In the non-relativistic limit, the SN equations and the GPN system are discussed. Ground state solutions of the SN 
  equations are shown to be stable, while radially exited states of the system were found to be unstable \cite{dissnewton}. In case of the GPN system, stable and unstable ground state solutions exist 
  \cite{harko2012,evolution2015,cana}, but the radially excited states seem to be unstable \cite{cooling2006}. However the investigation of excited states is still of interest. In case of the SN equations S.-J. 
  Sin had to use excited states with more than four nodes to get a satisfying approach of the observed rotation curves. Furthermore, L.A. Ure\~na-L\'opez and A. Bernal found that a superposition of the ground state 
  and excited states can be stable \cite{supstate}.  

  In this paper we continue the work of P.-H. Chavanis by calculating and analyzing the first three radially excited states (wave function solutions with a characteristic number of nodes) of the GPN system for 
  influence of the two external parameters: total mass and strength of the particle interaction. We restrict to non-rotating systems. 

  This paper is organized as follows. In section \ref{gppsystem} the GPN system is introduced as well as two of its limits: the TF limit and the non-interacting limit. In section \ref{numpro} the used numerical 
  procedure is explained and confirmed to work properly by comparing computed solutions with solutions published by other authors. In section \ref{solutions} the computed results of the first three radially excited 
  states are compared to the numerical results for the ground state. Ground state solutions were already discussed by P.-H. Chavanis \cite{cnum}, but are nevertheless computed in this paper as well for reasons of 
  completeness. In section \ref{stability} the stability properties of the found solutions are discussed. As already mentioned above, the radially excited states of the GPN system are found to be intrinsically 
  unstable by F. S. Guzm\'an and L.~A. Ure\~na-L\'opez \cite{cooling2006} by studying the time evolution of equilibrium solutions, while allowing a flow of particles out of the numerical domain. However the stability 
  of the GPN system was discussed only for a few different values of the parameter $g$, which indicates the strength of particle interactions. In this paper we use a different way to approach the stability question. 
  By applying arguments of the catastrophe theory we are able to give statements about a whole branch of possible configurations for a  self-gravitating BEC. Finally, in section \ref{conclusion}, we draw some 
  comments and conclusions.

\section{The Gross-Pitaevskii-Newton System}\label{gppsystem}

A BEC in its own gravitational field is described by the GPN system \cite{dmboehmer,cana} 
\begin{align}\begin{split}
i \hbar \frac{\partial \Psi(\vec{r},t)}{\partial t} = &-\frac{\hbar^2}{2m}\nabla^2\Psi(\vec{r},t) + m\Phi(\vec{r},t) \Psi(\vec{r},t) \\
& + g \left|\Psi(\vec{r},t)\right|^2\Psi(\vec{r},t) \end{split} \\
\Delta \Phi(\vec{r},t) = & \, \phantom{-} 4\pi G m\left|\Psi(\vec{r},t)\right|^2 \, .
\end{align}
With the self-interaction factor $g = 4\pi \hbar^2a/m$ short range interactions are taken into account. Here $a$ is the scattering length of the particles forming the BEC which can be chosen either positive (repulsive) or negative (attractive), and $m$ is the mass of each single particle which is part of the BEC.

The wave function $\Psi(\vec{r},t)$ satisfies the normalization condition
\begin{equation}
N = \int{ d^3 \vec{r}\;\left|\Psi(\vec{r},t)\right|^2},
\end{equation}
where $N$ is the total number of particles. $\left|\Psi(\vec{r},t)\right|^2$ corresponds to the particle density and $\rho(\vec{r})=m\left|\Psi(\vec{r},t)\right|^2$ gives the mass density. Thus, the total mass $M$ of the BEC can be computed with
\begin{equation}
M=mN=m\int d^3\vec{r}\,\left|\Psi(\vec{r},t)\right|^2 \, .
\label{gesamtmasse}
\end{equation}

We obtain the time-independent GPN system for stationary solutions of the form $\Psi(\vec{r},t)=\psi(\vec{r})\exp(-i E t / \hbar)$,
\begin{align}
E\psi(\vec{r}) &= -\frac{\hbar^2}{2m}\nabla^2\psi(\vec{r}) + m\Phi(\vec{r}) \psi(\vec{r}) + g \left|\psi(\vec{r})\right|^2\psi(\vec{r}) \label{t_ind_gpp1}\\
\Delta \Phi(\vec{r}) &= \phantom{-} 4\pi G m\left|\psi(\vec{r})\right|^2 \, , \label{t_ind_gpp2}
\end{align}
with $E$ being the eigenenergy of the GPN system. 

  \subsection{The energy functional}
    Now we introduce the funcional $E_{\text{tot}}[\psi]$ which is associated with the total energy of the GPN system:
    \begin{align} \begin{split}
      E_{\text{tot}}=\int d^3\vec{r} \biggl\{ &\frac{\hbar^2}{2m}\left|\nabla\psi (\vec{r}) \right|^2 +\frac{m}{2}\Phi(\vec{r})\left|\psi(\vec{r}) \right|^2 \\
      & +\frac{g}{2} \left|\psi (\vec{r}) \right|^4 \biggr\} \, .\end{split} 
    \end{align}
   Here $\Phi$ is a function of $\psi (\vec{r})$, according to \eqref{t_ind_gpp2}.
   An extremum of the total energy at fixed total mass $M$ is given by the variational principle 
   \begin{equation}
    \delta E_{\text{tot}}-\alpha\, \delta \left(m\int d^3\vec{r}\,\left|\Psi(\vec{r},t)\right|^2-M\right)=\delta I =0~~,
   \end{equation}
   where equation \eqref{gesamtmasse} is used. $\alpha$ is a Lagrange multiplier which takes into account the mass constraint and which can be interpreted as a chemical potential.
   The functional
   \begin{align} \begin{split}
    I =\int d^3\vec{r} \biggl\{ &\frac{\hbar^2}{2m}\left|\nabla\psi (\vec{r}) \right|^2 +\frac{m}{2}\Phi(\vec{r})\left|\psi(\vec{r}) \right|^2 \\
      & +\frac{g}{2} \left|\psi (\vec{r}) \right|^4 - \alpha\left|\psi (\vec{r}) \right|^2\biggr\} +\alpha M\, \end{split} \label{Wirkungsintegral}
   \end{align}
   is called the energy functional and its variation $\delta I$ is given by
   \begin{align}\begin{split}
    \delta I =& \int d^3\vec{r} \biggl\{-\frac{\hbar^2}{2m}\left(\Delta \psi \delta\psi^* + \Delta\psi^*\delta\psi \right) \\
    & + (m\Phi +g\left|\psi\right|^2 -\alpha ) \left(\psi \delta\psi^* + \psi^*\delta\psi \right)\biggr\}~~. \end{split}
    \label{varwirkung}
   \end{align}
   Since $\Phi$ depends on $\psi$, one has to variate $\Phi$ in \eqref{Wirkungsintegral} as well. This results in the loss of a $1/2$ factor from \eqref{Wirkungsintegral} to \eqref{varwirkung} at the $\Phi$-term.
   By identifying $\alpha$ as eigenenergy $E$, solving $ \delta I/\delta \psi =0$ or $\delta I/\delta\psi^* = 0$ results in the time-independent GPN system \eqref{t_ind_gpp1} and \eqref{t_ind_gpp2}. 
   Wave functions, which correspond to extrema of the energy functional are therefore solutions to the GPN system.

\subsection{Limiting cases of the GPN system} 
We now describe two threshold regions of the solution manifold of the GPN system which we will need later. 
\begin{itemize}
\item[(i)] In the \textit{non-interacting limit} 
the self-interaction term in \eqref{t_ind_gpp1} can be neglected
\begin{equation}
g \left|\psi(\vec{r})\right|^2\psi(\vec{r}) \longrightarrow 0 \, .
\end{equation}
Then the GPN system reduces to the SN system. 
The non-interacting limit is obtained for $g=0$ or very small total masses $M$ (i.e. $\left|\psi(r)\right|^2 \rightarrow 0$).
  \item[(ii)] The \textit{Thomas-Fermi limit} (TF limit) is characterized by a large number of particles so that the kinetic term in \eqref{t_ind_gpp1}, $-\frac{\hbar^2}{2m}\nabla^2\psi(\vec{r})$, can be neglected. This limit was studied in \cite{dmboehmer}.
  Due to the dominant resulting repulsive interactions (solutions in the TF-limit can only be found for $g>0$) the self-gravitating BEC is prevented from gravitational collapsing. In the TF limit we obtain an approximate 
  solution for the ground state of the GPN system. This is due to the fact that in the TF limit solutions can only be found if there exists a repulsive 
force $F_{\text{si}}(\vec{r})$ for all $\vec{r}$. This force is caused by the self-interaction term $g \left|\psi(\vec{r})\right|^2\psi(\vec{r})$ and is given by
\begin{equation}
F_{\text{si}}(\vec{r})=\frac{g}{m}\nabla\rho (\vec{r}) \, .
\end{equation}  
A repulsive force $F_{\text{si}} (\vec{r}) <0  \, \forall \vec{r}$ can be found only for the ground state for which the condition $\nabla\rho(\vec{r})<0 \; \forall \vec{r}$ is satisfied. Excited states do not fulfill this condition. 
	    

The density profile $\rho(r)$ of a spherical symmetric GPN system in the TF limit is given by \cite{cana}
\begin{equation}
\rho(r) = \frac{\rho_0 R_{\text{TF}}}{\pi r} \sin\left(\frac{\pi r}{R_{\text{TF}}}\right)\text{, for }r\leq R_{\text{TF}} \, ,\label{rho}
\end{equation}
where $\rho_0$ is the central density at $r=0$, and $R_{\text{TF}}=\pi(a \hbar^2/Gm^3)^{1/2}$ is the radius of the BEC where $\rho(r)$ becomes zero. The eigenenergy $E$ is determined as 
\begin{equation}
E = -\frac{GMm}{R_{\text{TF}}}~~.\label{energietf}
\end{equation}  
\end{itemize}

\section{Solving the GPN system}\label{numpro}

As we are interested in radially exited states only, we choose a spherically symmetric ansatz for the wave function, $\psi(\vec{r}) = f(r)$. In the following we use $E=\hbar\omega$. 

\subsection{Rescaling} 

For solving the GPN system it is convenient to rescale the variables and make them dimensionless. The introduction of natural length and energy units $\hat{r}=\frac{2m^3G}{\hbar^2}\,r$ and $\hat{\omega}= \frac{\hbar^3}{2G^2m^5}\,\omega$ results in the rescaling of the other variables. With a further dimensionless scaling factor $\lambda$ with $\lambda \in \left\{ \mathbb{R}\,|\,\lambda>0\right\}$ we obtain
\begin{equation}
\begin{aligned}
g &= \pi\frac{\hbar^4}{2m^4G\lambda^2}\hat{g}\, ,    & f(r) &= \frac{1}{\sqrt{4\pi}}\left(\frac{2m^3G}{\hbar^2}\right)^{3/2}\lambda^2\hat{f}(r)\, ,\\
\omega &= 2\frac{G^2m^5}{\hbar^3}\lambda^2\hat{\omega}\, , & \Phi(r) &= \frac{2G^2m^4}{\hbar^2}\lambda^2\hat{\Phi}(r)\, ,\\
r &= \frac{\hbar^2}{2m^3G\lambda}\hat{r}\,  ,  & M  &=\lambda m \hat{M} = \lambda m\int d\hat{r}\, \hat{r}^2 \hat{f}(\hat{r})^2\, .\\
\end{aligned}\label{skalierung}
\end{equation}
This rescaling results in the following dimensionless time-independent GPN system
\begin{align}\begin{split}
\frac{d^2 \hat{f}(\hat{r})}{d \hat{r^2}} &= \hat{\phi}(\hat{r})\hat{f}(\hat{r})-\hat{\omega} \hat{f}(\hat{r}) -\frac{2}{\hat{r}}\frac{d \hat{f}(\hat{r})}{d \hat{r}} + \hat{g} \hat{f}(\hat{r})^3 \, , \\
\frac{d^2 \hat{\phi}(\hat{r})}{d \hat{r}^2} &= \hat{f}(\hat{r})^2-\frac{2}{\hat{r}}\frac{d \hat{\phi}(\hat{r})}{d \hat{r}} \, . \end{split} \label{dlgpn2}
\end{align} 
The scaling factor $\lambda$ does not appear in \eqref{dlgpn2} showing that $\lambda$ leaves \eqref{dlgpn2} form invariant. This characteristic feature can be used to obtain solutions for different values of $\hat{g}$ and $\hat{M}$ directly from already computed solutions, by changing the value of $\lambda$.
  
In the following, the dimensionless GPN system \eqref{dlgpn2} will be solved numerically by using the FORTRAN program COLSYS.

\subsection{The numerical procedure} 

COLSYS can be used for numerically solving mixed-order problems of systems of ordinary differential equation (ODE) with given boundary conditions. (More information can be found in \cite{ascher}.)
In order to enable the computation of the eigenenergy $\hat{\omega}$ of the dimensionless GPN system \eqref{dlgpn2} an additional differential equation has to be added
\begin{equation}
\frac{d^2\hat{\omega}}{d \hat{r}^2}=0.
\end{equation}
This equation completes the set of ODEs.

The boundary conditions for the system of ODEs are chosen as follows:\\
\begin{align}
& \text{a)} & \hat f^\prime(0) & = 0\,, \qquad \text{d)} & \hat f(R) & = 0\,, \nonumber\\
& \text{b)} & \hat\phi^\prime(0) & = 0\,, \qquad \text{e)} & \hat\phi(R) & = - R \hat\phi^\prime(R)\,, \\
& \text{c)} & \hat\omega^\prime(R) & = 0\,, \qquad \text{f)} & \hat f(0) & = f_c \;\; \text{or} \;\; \hat\phi^\prime(R) = \frac{\hat M}{R^2}\,, \nonumber
\end{align}
where the prime denotes $d/d\hat{r}$. $R$ represents a fixed and very large value of $\hat{r}$ which will be determined later. It is used to fix the behavior of the solution far away from the origin. The boundary condition f) can be used to preset either a value for $\hat{f}(0)$ or the total mass $\hat{M}$ of the BEC.

\begingroup
\squeezetable
\begin{table}[b]
\begin{center}
\begin{tabular}{ c||c|c|c|c}
state  & $\hat{\omega}$ & $2\cdot \hat{\omega}$ & $\omega_H$  in \cite{dissnewton} & $\Delta_{2\hat{\omega},\omega_H}$ \\ \hline\hline
ground state & $-0.081384604$ & $-0.162769208$ & $-0.162769291$ & $5 \cdot 10^{-7}$ \\
1. ex. state & $-0.015398269$ & $-0.030796537$ & $-0.030796561$ & $7 \cdot 10^{-7}$ \\
2. ex. state & $-0.006263051$ & $-0.012526101$ & $-0.012526108$ & $6 \cdot 10^{-7}$ \\
3. ex. state & $-0.003373660$ & $-0.006747320$ & $-0.006747330$ & $1 \cdot 10^{-6}$ \\
\end{tabular}
\end{center}
\caption{Computed values of the eigenenergy $\hat{\omega}$ of the ground state ($k=0$) and the first three radially excited states ($k=1,2,3$) for $\hat{g}=0,\hat{M}=1$ compared with the values of the eigenenergy 
$\omega_H$ computed in \cite{dissnewton}. Deviations $\Delta_{2\hat{\omega},\omega_H}$ of $2\hat{\omega}$ from $\omega_H$ do not exceed $10^{-6}$.}
\label{wvergleich}
\end{table}
\endgroup
We start our computations with solutions (ground state and the first three radial excitations) characterized by $\hat{g} = 0$ and $\hat{M} = 1$. The obtained solutions are compared to those computed 
in \cite{dissnewton} (in fact, by our method we were able to reproduce the solutions).
Branches of ground state solutions and branches of solutions of the first three radially excited states, respectively, can be computed by slowly changing the value of $\hat{g}$ or of $\hat{M}$. 
A full analysis of the influence of the parameters $\hat{g}$ and $\hat{M}$ on the GPN system needs to either set a fixed value of $\hat{M}$ and vary $\hat{g}$ or to vary 
$\hat{M}$ for a fixed $\hat{g}>0$ and a fixed $\hat{g}<0$. Solutions for other combinations of $(\hat{g},\hat{M})$ are then obtained from these computed solutions by the variation of the value of the 
scaling factor $\lambda$. 

For a comparison of the obtained solutions we match the values of the computed eigenenergy $\hat{\omega} = \hat{E}/ \hbar$ with the appropriate values $\omega_H$ calculated in \cite{dissnewton}.
Since in \cite{dissnewton} $N,\hbar,G,m,\lambda$ are set equal to $1$, $\omega_H$ has to coincide with $2\hat{\omega}$ (see scaling condition for $\omega$, equation \eqref{skalierung}). Table \ref{wvergleich} 
summarizes the comparison of our results with those in \cite{dissnewton}. Deviations are at most of order of $10^{-6}$. Thus, our ansatz and software seems to work correctly. Furthermore the shapes of the computed ground state wave function and the corresponding gravitational potential for $\hat{g}=0$ and $\hat{M}=1$ are compared with the results obtained by R. Ruffini et al. in \cite{ruffini}. The calculated functions perfectly coincide what further supports our conclusion.


\section{Solutions}\label{solutions}
\begin{figure}[htp]
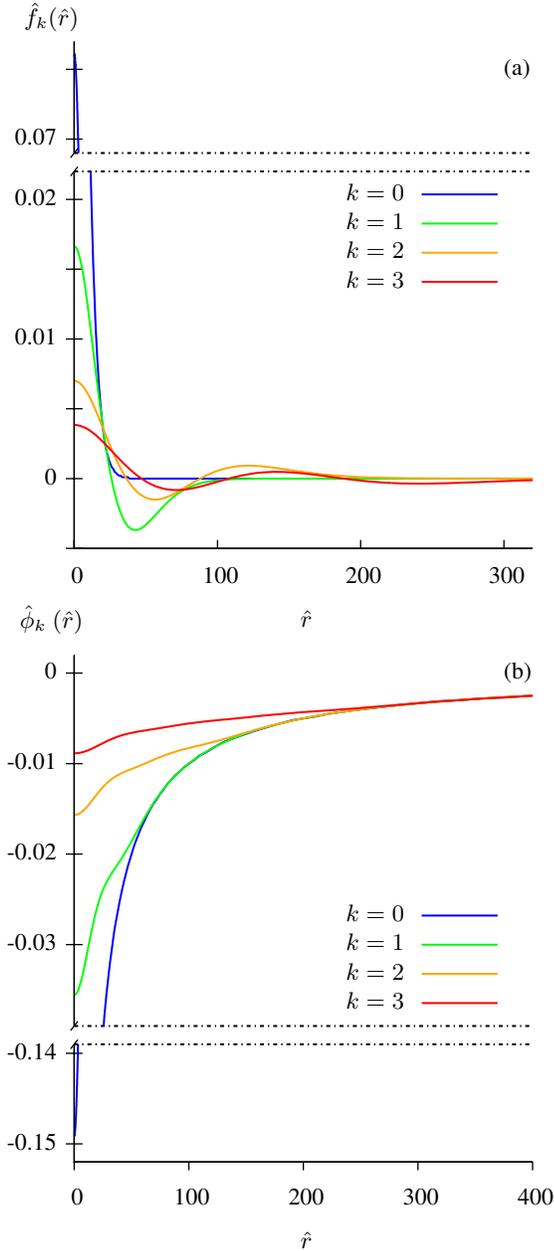

\small
\centering
\input{Figures/f_g1_m1}
\input{Figures/phi_g1_m1}
\caption{(a) Computed solutions for the wave function and (b) computed solutions for the gravitational potential of the ground state ($k=0$) and the first three radially excited states ($k=1,2,3$) for $\hat{g}=\hat{M}=1$.}
\normalsize\label{wave}
\end{figure}

Typical profiles of the obtained wave functions and their corresponding gravitational potentials of the GPN system are shown exemplarily in Fig.~\ref{wave}. The number of nodes of a wave function is denoted $k$, the wave 
function itself is denoted $\hat{f}_k$ and the corresponding gravitational potential is denoted $\hat{\phi}_k$. For the ground state we have $k=0$ and the first three radially excited states are indicated by $k=1,2,3$. 
The solutions in Fig.~\ref{wave} correspond to $\hat{g}=\hat{M}=1$. We observe that for larger $k$ the maximum of $\hat{f}_k(\hat{r})$ at the origin decreases, and that $\hat{f}_k(\hat{r})$ more slowly approaches the abscissa.
Thus, radially excited solutions of self-gravitating BECs have lower density in the center and a larger extension. Nodes of the radially excited states create ``bulges'' in their associated gravitational potentials. 

\subsection{The total mass -- eigenenergy relation}    
\begin{figure}[floatfix]
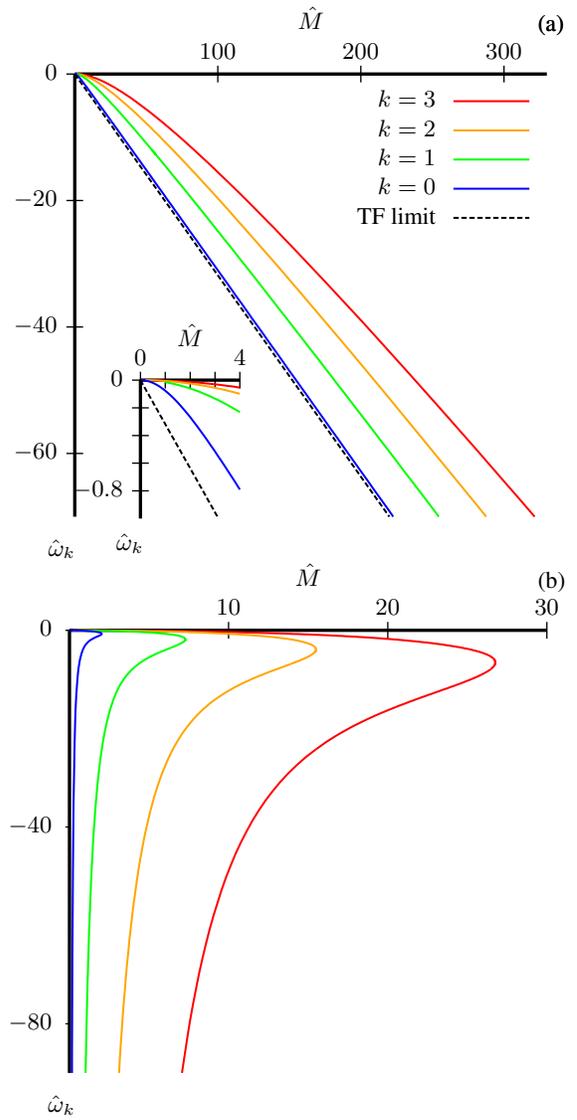

\small
\centering
\input{Figures/w_ueber_m_g1}
\input{Figures/w_ueber_m_gn1}
\caption{The eigenenergy $\hat{\omega}_k$ is plotted as a function of the total mass $\hat{M}$ for (a) $\hat{g}=1$, representing $\hat{g}>0$ and (b) $\hat{g}=-1$, representing $\hat{g}<0$. For the ground state ($k =0$) and 
	 the radially excited states ($k=1,2,3$) the course of $\hat{\omega}_k(\hat{M})$ shows the same qualitative shape. In case of a fixed $\hat{g}>0$ and sufficiently high values of 
	 $\hat{M}$, $\hat{\omega}_0(\hat{M})$ closely approaches the course of the TF limit which is described by $\hat{ \omega}=-\hat{r}/(\pi \sqrt{\hat{g}})$. In case of a fixed $\hat{g}<0$ a maximum value of the total 
	 mass is found, $\hat{M}_{\text{max},k}$ (for $k = 0,1,2,3$). The value of $\hat{M}_{\text{max},k}$ increases the higher the system is radially excited. In case of $\hat{M}<\hat{M}_{\text{max},k}$ two equilibrium 
	 solutions with the same number of nodes and for a fixed pair of parameters $(\hat{g},\hat{M})$ can be found.}
\normalsize \label{wum}
\end{figure}
 
\begin{figure}[floatfix]
\small
\begin{center}
\input{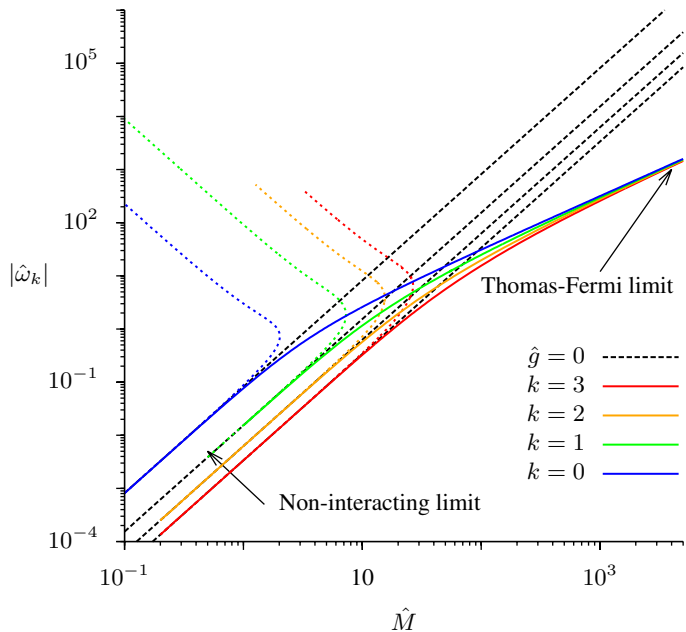}
\caption{The absolute value of the eigenenergy $\left|\hat{\omega}_k\right|$ is shown as a function of the total mass $\hat{M}$ for $\hat{g}=1,0,-1$ in a log-log plot.  Dashed colored lines correspond to $\hat{g}=-1$, 
	 solid lines correspond to $\hat{g}=1$, and dashed black lines correspond to $\hat{g}=0$. Every case is plotted for the ground state ($k = 0$) and the first three radially excited states ($k=1,2,3$). For a 
	 small value of the total mass all graphs follow the relation $\left|\hat{\omega}_k\right|\propto \hat{M}^2$ and the non-interacting limit $(\hat{g}=0)$ turns out to be a sufficient description of the system. 
	 For a rising value of $\hat{M}$ all graphs with a fixed $\hat{g}>0$ slowly reach a linear dependence on $\hat{M}$ with the same slope. Here $\hat{\omega}_k(\hat{M})$ follows the course of the TF limit except 
	 for a constant offset.}  \label{wumlog}
\end{center}
\normalsize
\end{figure}

    In Fig.~\ref{wum} the eigenenergy $\hat{\omega}_k$ of a wavefunction solution with $k$ nodes is plotted as a function of the total mass $\hat{M}$ (a) for a fixed value $\hat{g}>0$ and (b) for a fixed value $\hat{g}<0$, 
    respectively. These two plots are combined in a log-log plot in Fig.~\ref{wumlog}. For $\hat{g}>0$ and $\hat{g}<0$ the behavior of $\hat{\omega}_k(\hat{M})$ shows the same qualitative shape for the ground state ($k=0$) 
    and the first three radially excited states ($k=1,2,3$).
    For small masses $\hat{M}$ and $\hat{g} \not= 0$ the function $\hat{\omega}_k(\hat{M})$ looks like $\hat{\omega}(\hat{M})$ in the non-interacting limit $\hat{g} = 0$.
    This behavior is described by
    \begin{equation}
      \hat{\omega}_k = \frac{\hat{\omega}_{k,c}}{\hat{M}_c^2}\hat{M}^2~~. \label{energieni}
    \end{equation}
    $\hat{\omega}_{k,c}$ is the computed eigenenergy of a wave function solution with $n$ nodes for $\hat{M}=\hat{M}_c$. $\hat{M}_c$ can be chosen arbitralily. The relation \eqref{energieni} can be deduced from the 
    rescaling relation of $\hat{M}$ and $\hat{g}$ \eqref{skalierung}.

    For a fixed value $\hat{g}>0$, $\hat{\omega}_0(\hat{M})$ of the set of ground state solutions closely approaches the course of $\hat{\omega}(\hat{M})$ in the TF limit, as is shown in Fig.~\ref{wum}. The course of 
    $\hat{\omega}(\hat{M})$ in the TF limit is computed by using equation \eqref{energietf} and by converting it into the rescaled variables, see \eqref{skalierung}. As predicted by \eqref{energietf}, the eigenenergy in 
    the TF limit linearly depends on the total mass. 

    For higher values of the total mass the course of $\hat{\omega}_k(\hat{M})$ approaches a linear behavior for all $k=0,1,2,3$ with the slope of $\hat{\omega}$ in the TF limit, see Fig.~\ref{wumlog}. 
    Here $\hat{\omega}_k(\hat{M})$ follows the TF limit except for a constant offset. This constant offset increases with the radial excitation, see Fig.~\ref{wum}.

    For a fixed value $\hat{g}<0$ the sets of solutions of the ground state and the first three radially excited states possess a maximum value of the total mass, $\hat{M}_{\text{max},k}$. For chosen values of 
    the total mass $\hat{M}$ with $\hat{M}> \hat{M}_{\text{max},k}$ no equilibrium solution with the respective excitation can be found, see Figs.~\ref{wum} and \ref{wumlog}. Furthermore, we find 
    $\hat{M}_{\text{max},k}< \hat{M}_{\text{max},(k+1)}$. Thus, for values of $\hat{M} > \hat{M}_{\text{max},k}$ of a specific set of solutions (which is characterized by a fixed value of $\hat{g}$ and a fixed 
    node number $k$) it is only possible to find equilibrium solutions (with the same value for $\hat{g}$) with an increased number of nodes. For every total mass $\hat{M}$ smaller than $\hat{M}_{\text{max},k}$ 
    two solutions with $k$ nodes for the same value of $\hat{g}$ can be found. Thus, for the set of solutions with $k$ nodes and a fixed $\hat{g}$ we obtain two branches of solutions for $\hat{M} < \hat{M}_{\text{max},k}$. 
    The branch of solutions with higher values of the eigenenergy $\hat{\omega}_k$ is considered to be stable, the other branch contains solutions which are treated unstable. 
    This issue is explained in more detail in the following section.
    
    For sufficiently small values of $\hat{M}$ the eigenenergy $\hat{\omega}_k$ of the unstable branch turns to be proportional to $-\frac{1}{\hat{M}^2}$, see Fig.~\ref{wumlog}).
 
\subsection{The self-interaction parameter -- eigenenergy relation}

\begin{figure}[t!] 
\small
\begin{center}
\input{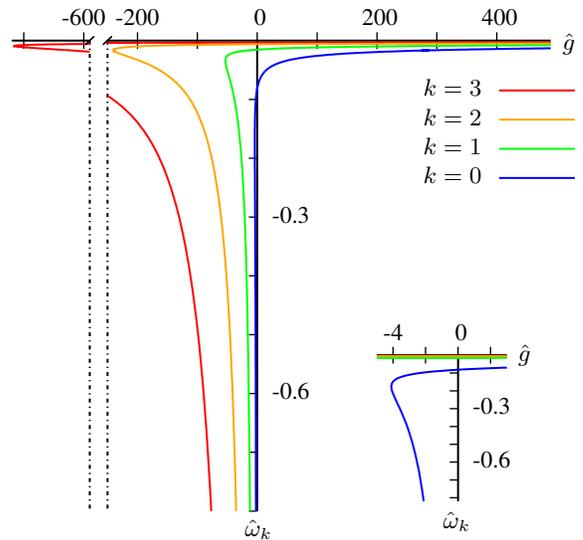}
\caption{The eigenenergy $\hat{\omega}_k$ is plotted as a function of the self-interaction factor $\hat{g}$ for $\hat{M}=1$. For the ground state ($k=0$) and the first three radially excited states ($k=1,2,3$) $\hat{\omega}_k(\hat{g})$ shows the 
	 same qualitative shape. It exists a minimal value of the self-interaction factor $\hat{g}_{\text{min},k}<0$ for each set of solutions with $k$ nodes. The value of $\hat{g}_{\text{min},k}$ decreases for larger $k$. For a fixed 
	 $\hat{g}_{\text{min},k}<\hat{g}<0$ it is possible to compute two solutions with the same number of nodes and the same total mass $\hat{M}$.} \label{wug}
\end{center}
\normalsize
\end{figure}

In Fig.~\ref{wug} the eigenenergy $\hat{\omega}_k$ is plotted as a function of the self-interaction parameter $\hat{g}$ for a fixed value of the total mass $\hat{M}$. The function $\hat{\omega}_k(\hat{g})$ shows the same 
qualitative shape for the ground state ($k=0$) as well as for the first three radially excited states ($k=1,2,3$). There exists a minimum value $\hat{g}_{\text{min},k}<0$ of the self-interaction parameter for sets of 
solutions with $k$ nodes and a fixed value of $\hat{M}$. The value $\hat{g}_{\text{min},k}$ is smaller the higher the system is radially excited. For a fixed 
$\hat{g}_{\text{min},k}<\hat{g}<0$ it is possible to compute two solutions with the same total mass $\hat{M}$ and the same number of nodes $k'$, if $k'\geq k$.

\subsection{Thomas-Fermi limit of high total masses}
\begin{figure*}[floatfix]
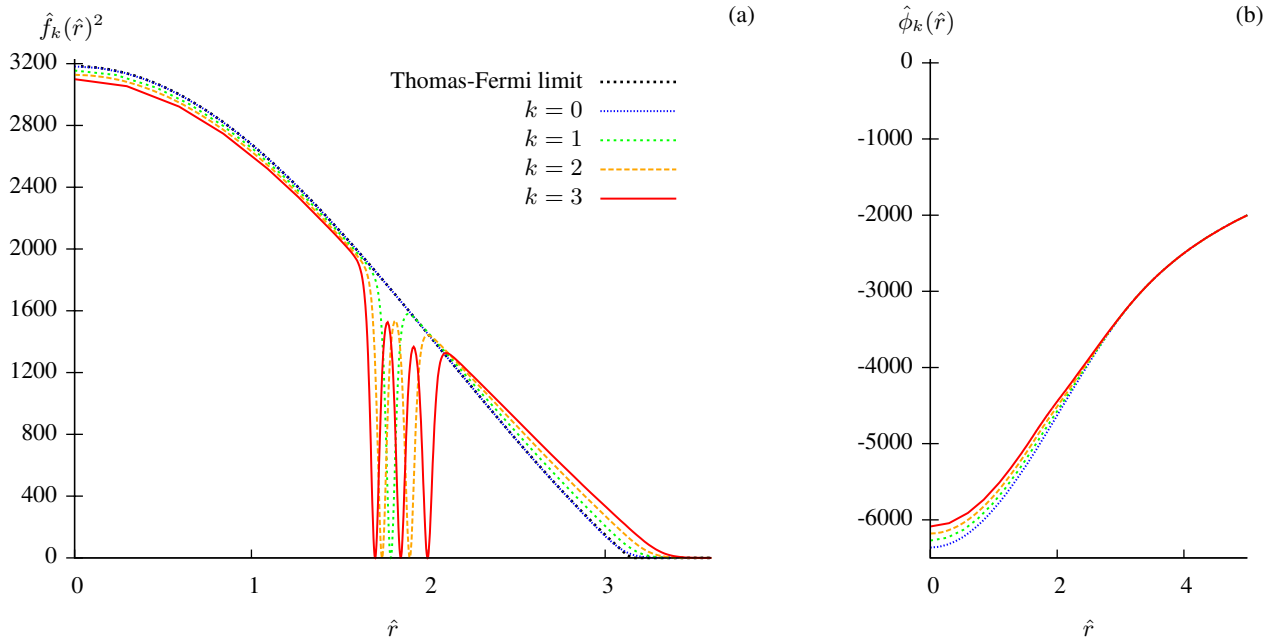

\centering
\begin{minipage}[b]{0.495 \textwidth}
\centering
\input{Figures/rho_g1_m10000}
\end{minipage}
\begin{minipage}[b]{0.495 \textwidth}
\centering
\input{Figures/phi_g1_m10000}
\end{minipage}
\caption{(a) Particle density profile $\hat{f}_k(\hat{r})^2$ and (b) the appropriate gravitational potentials $\hat{\phi}_k(\hat{r})$ of the ground state ($k=0$) and the first three radially excited 
	 states $(k = 1,2,3)$ for $\hat{g}=1$ and for a high total mass ($\hat{M}=10^4$). $\hat{f}(\hat{r})^2$ in the TF limit satisfies $\hat{f}(\hat{r})^2=\pi \hat{M}/(\hat{g}\hat{r})\sin(\hat{r}/\sqrt{\hat{g}})$ 
	 and is plotted in black. The profile nearly coincides with $\hat{f}_0(\hat{r})^2$. $\hat{f}_k(\hat{r})^2$ for $k=1,2,3$ seem to approach $\hat{f}(\hat{r})^2$ of the TF limit as well, disturbed though, by gaps within their courses, 
	 which occur due to number of nodes of the wave function solutions.}\label{bigmass}
\end{figure*}

For a positive value of $\hat{g}$ and a sufficiently large total mass $\hat{M}$ the behavior of $\hat{\omega_k}(\hat{M})$ is well approximated by its TF limit. For analyzing this regime of sets of solutions in case of ground state solutions and their 
first three radial excitations we compute the particle densities $\hat{f}_k(\hat{r})^2$ and the appropriate gravitational potentials for $\hat{M}=10^4$.
Furthermore, we calculate $\hat{f}(\hat{r})^2$ of the corresponding TF-limit by using equation \eqref{rho} and by converting it into the rescaled variables \eqref{skalierung} (see Fig.~\ref{bigmass}).

    While the particle density profile of the ground state is well described with the particle density profile of the TF limit, the profiles of the radially excited states seem to approach that of the TF limit as 
    well except for small gaps. These gaps occur owing to the nodes of the wave functions. In contrast to small $\hat{M}$, for large $\hat{M}$ these disturbing gaps in the particle density profiles appear much more 
    abrupt, see Fig.~\ref{bigmass}. Furthermore, for large $\hat{M}$ the bulges in the shape of the gravitational potentials nearly disappear, see Fig.~\ref{bigmass}. 

    Comparable results for the behavior of excited states of a BEC, but by assuming the BEC to be located in a harmonic trap, are computed in \cite{zeli} for a large self-interaction. Note that for the used scaling \eqref{skalierung}  solutions of the GPN system at large total masses can be calculated from solutions for large self-interaction factors. This is possible, since one 
    can calculate solutions for different parameters $\hat{g}_a$,$\hat{M}_a$ and $\hat{g}_b$,$\hat{M}_b$ from each other as long as the equation $\hat{g}_a\hat{M}_a^2=\hat{g}_b\hat{M}_b^2$ holds. Thus, solutions 
    characterized by large values of $\hat{g}$ correspond to solutions characterized by large values of $\hat{M}$. In \cite{zeli} the behavior of the wave functions is referred to a strong repulsion between 
    particles, which leads to a spatial distribution as even as possible. A sharper decrease of the wave function at the node locations allows a more even particle distribution.

    In case of the present GPN system, sharp changes of the wave function of radially excited states at narrow ranges around the nodes lead to comparably small changes at the other parts of the wave function.
    Here the kinetic term can be neglected and, thus, the course of the wave function again approaches the TF limit. Therefore, radially excited states also approach the TF limit, except for a small range where 
    gaps are occurring. The extension of the gaps shrinks with a rising total mass. 

\section{Stability analysis}\label{stability}

For the stability analysis of solutions of the dimensionless time-independent GPN system \eqref{dlgpn2} arguments of catastrophe theory are applied. Likewise, this method was used in several publications 
on the stability of boson stars (e.g. \cite{bstar91,sak,bstar11,bstar12}). A general introduction to and applications of catastrophe theory can be found in, e.g., \cite{saun,dom,pos}. Here we proceed similarly 
to the procedure employed by N. Sakai et al. \cite{sak}. 

With the help of catastrophe theory it is possible to discuss the critical points of a nonlinear system. If such a system is described by a potential then its solutions are extrema of this potential and its critical 
points. In the present case we assume the appropriate energy functional \eqref{Wirkungsintegral} is the characterizing potential. Solutions of the dimensionless and time-independent GPN system correspond to extrema 
of this energy functional.

The stability of the solutions will be discussed for elements of the function space $D_k$ containing the set of spherically symmetric functions characterized by $k$ nodes. The stability of the ground state and the 
radially excited states will be discussed separately. 

\subsection{Procedure to apply catastrophe theory}
\begin{figure*}[ht] 
\small
 \begin{center}
\input{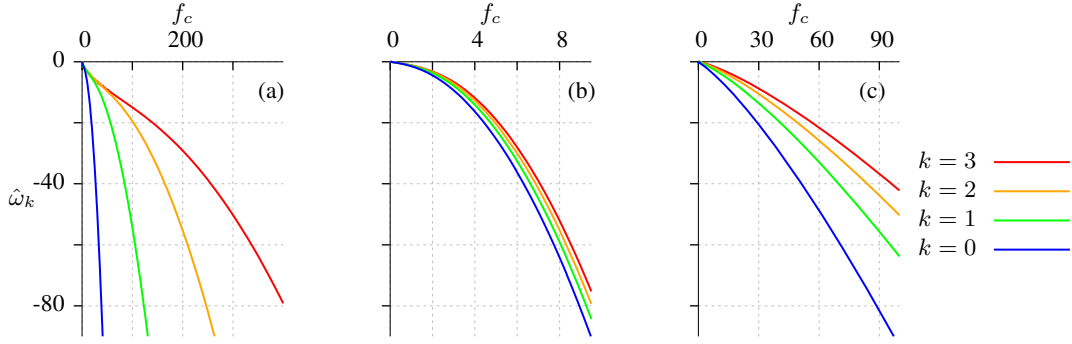}
\caption{The eigenenergy $\hat{\omega}_k$ is plotted as a function of $f_c$ with (a) a fixed $\hat{g}=-1$, (b) a fixed $\hat{g}=1$ and (c) a fixed $\hat{M}=1$. The plots indicate that $\hat{\omega}_k$ is a strictly decreasing function of 
	 $f_c$.} \label{fw}
\end{center}
\normalsize
\end{figure*}
As a first step we choose suitable control parameters and a suitable behavior variable of the system. In our case the behavior of the characterizing potential $\mathcal{I}$ derived from the functional $I$ depends on $\hat{M}$ and $\hat{g}$. Since they can be given by hand, $\hat{M}$ and $\hat{g}$ act as control parameters of the system. The characterizing potential is described as a function of the behavior variable.

To introduce a behavior variable $x$ for our system we use a one parameter family of perturbed functions $\hat{f}_{k,x}(\hat{r})$ near the equilibrium solution $\hat{f}_k(\hat{r})$. For these functions the energy 
functional $I$ \eqref{Wirkungsintegral} can then be regarded as a function $\mathcal{I}(x) := I[\hat{f}_{k,x}]$ of $x$. This allows us to introduce the critical points of the characterizing potential $\mathcal{I}(x)$. They are defined as points $x_c$, at which the derivation $d\mathcal{I}(x)/dx$ vanishes. In the case that the functional $I$ is calculated with $\hat{f}_{k,x}(\hat{r})$, then the equation $dI [\hat{f}_{k,x}] /dx = (\delta I/ \delta \hat{f}_{k,x}) d\hat{f}_{k,x}/dx$ holds. Then we have
\begin{equation}
\frac{\delta I[\hat{f}_{k,x}]}{\delta \hat{f}_{k,x}} = 0\quad \Rightarrow\quad \frac{d\mathcal{I}(x)}{dx} =0~~.
\label{critpoint}
\end{equation}
From this it follows that $\delta I/\delta \hat{f}_{k,x} =0$ is satisfied if $x=x_c$. Then $\hat{f}_{k,x_c}$ is exactly the equilibrium solution $\hat{f}_k$. Therefore critical points $x_c$ of $\mathcal{I}(x)$ 
represent solutions of the GPN system. 

It is suitable to choose $f_c=\hat{f}(0)= x$ as the behavior variable, since it describes the system uniquely on $D_k$ for varying values of the control parameters $\hat{g}$ and $\hat{M}$. This is meant in the sense that one finds for every 
value of $f_c$ exactly one solution for the wave function on the function space $D_k$ of the system. The corresponding family of perturbed functions can be specified by the relation 
\begin{equation}
\hat{M}=\int{ d\hat{r}\,\hat{r}^2\hat{f}_{k,f_c}(\hat{r})^2} \, . 
\end{equation}

In the present case it is possible to use the eigenenergy $\hat{\omega}$ as behavior variable instead of $f_c$. The GPN system has already been studied regarding $\hat{\omega}$ in the previous paragraphs. It also turns out that 
$\hat{\omega}_k(f_c)$ is a strictly decreasing function of $f_c$ for all $k$ (see Fig.~\ref{fw}). Thus, the behavior of the system does not change by changing its behavior variable from $f_c$ to $\hat{\omega}$ in any case.

Now the stability analysis procedure will be introduced. For each state one has to: 
\begin{enumerate}
\item Compute the equilibrium space $\mathcal{M}_k=\{\hat{\omega}_k,\hat{M},\hat{g}\}$, that contains all critical points of $I[\hat{f}_{k,\hat{\omega}}]$, $\hat{f}_{k,\hat{\omega}}\in D_k$. Each point in $\mathcal{M}_k$ represents a configuration 
      of the self-gravitating BEC. 
\item Determine the degenerate points or turning points of the potential, which satisfy $\frac{\partial \hat{M}}{\partial \hat{\omega}_k}=0,\; \frac{\partial \hat{g}}{\partial \hat{\omega}_k}=0$. These points are centers of catastrophes and stability 
      changes occur here. The set of all degenerate points in $\mathcal{M}_k$ is called catastrophe map 
      $\Sigma_k=\{\hat{\omega}_k,\hat{M},\hat{g}\left| \frac{\partial \hat{M}}{\partial \hat{\omega}_k}=0,\; \frac{\partial \hat{g}}{\partial \hat{\omega}_k}=0\right.\}$.
\item Compute the energy functional $I$ for equilibrium solutions $\hat{f}_k$ in the neighborhood around a degenerate point $p\in\Sigma_k$ in order to be able to assign stability properties to each point in $\mathcal{M}_k$.
\item Draw the bifurcation set $\xi_k$ in the control space $\mathcal{C}=\{\hat{M},\hat{g}\}$ and identify the regions of stability and instability, respectively. $\xi_k$ is a projection of $\Sigma_k$ onto the control space.
\end{enumerate}

\subsection{Domain of stability of solutions}
\begin{figure*}[t!]
\vspace{-1.cm}
\small
\begin{center}
\input{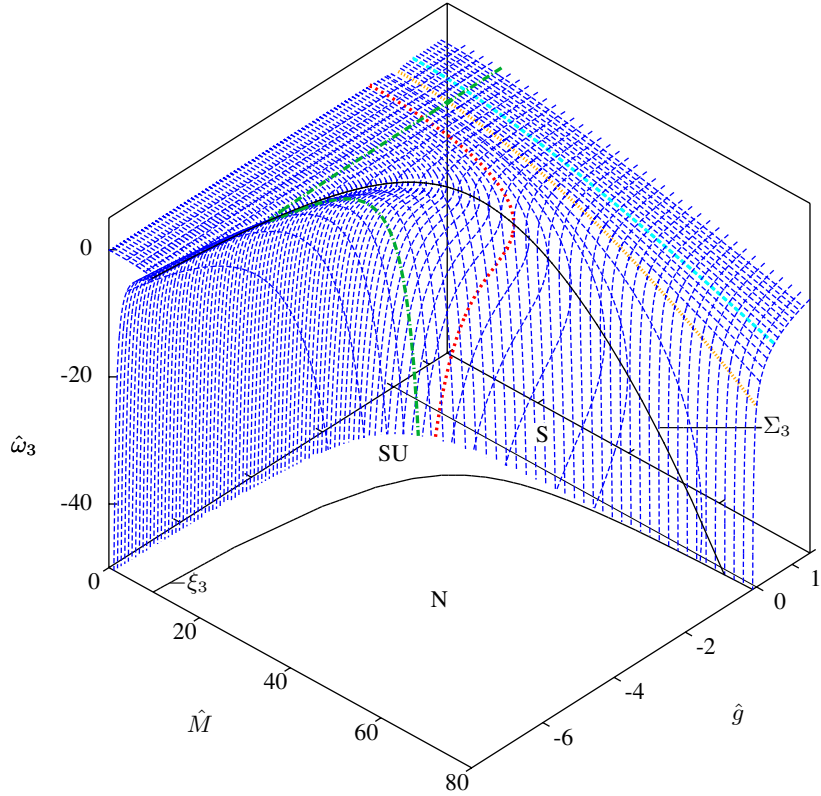}
\caption{The equilibrium space $\mathcal{M}_3$ contains all critical points $(\hat{M},\hat{g},\hat{\omega}_3)$ belonging to solutions of the third radially excited state of the GPN system. In order to enable a 
	 better overview, $\hat{\omega}_3$ is highlighted as a function of $\hat{g}$ for a fixed total mass $\hat{M}$ (green) and as a function of $\hat{M}$ for a fixed self-interaction factor $\hat{g}<0$ 
	 (red), $\hat{g}=0$ (orange) and  $\hat{g}>0$ (cyan). The catastrophe set $\Sigma_3$ and the bifurcation set $\xi_3$ are plotted in black. $\xi_3$ is a projection of $\Sigma_3$ onto the control plane $\mathcal{C}$. 
	 Together with the ($\hat{g}=0$)-axis it divides $\mathcal{C}$ in three regions: S, SU and N. For a fixed pair of control parameters $(\hat{M},\hat{g}) \in \mathcal{C}$ one finds one stable solution, one stable 
	 and one unstable solution or no equilibrium solution, respectively, of the third radially excited state. }
\label{ggm}
\end{center}
\normalsize
\vspace{-0.7cm}
\end{figure*}

Fig.~\ref{ggm} shows the equilibrium space $\mathcal{M}_3$. This figure is a combination of Figs.~\ref{wum} and \ref{wug} for $k=3$ where a qualitatively equal shape of the plotted functions for all $k$ was found. Due to this, 
the shape of the respective equilibrium spaces $\mathcal{M}_k$ do not qualitatively differ, too, and we exemplarily show the equilibrium space $\mathcal{M}_3$ only. Each point contained in $\mathcal{M}_3$ represents a solution with node number $k=3$. 
The catastrophe map $\Sigma_k$ is plotted as well. It divides the equilibrium space surface in an upper part and a lower part.

\begin{figure}[floatfix] 
\small
\begin{center}
\input{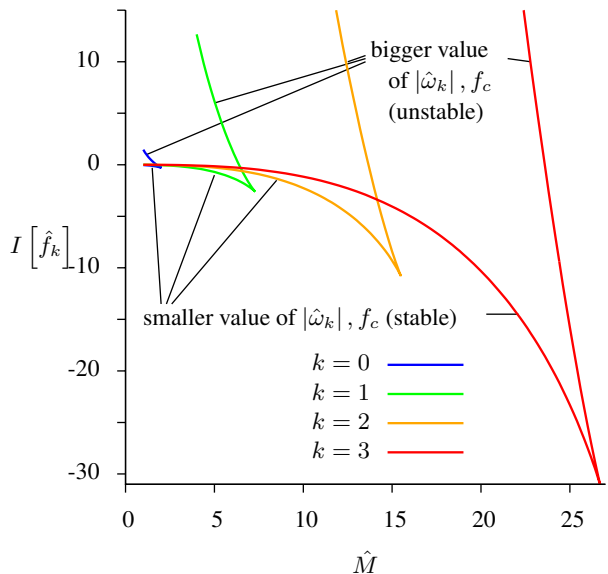}
\caption{The values of the energy functional $I$ of the equilibrium solutions is shown as a function of $\hat{M}$ for $\hat{g}=-1$. Here the lower branches for $k =0,1,2,3 $ are considered to correspond to stable solutions of the system; 
	 the upper branches correspond to unstable solutions of the system. For all states the course of $I$ shows the same qualitative shape.} 
\label{etotum}
\end{center}
\normalsize
\end{figure} 

In Fig.~\ref{etotum} the values of energy functional $I$ of the equilibrium solutions is plotted as a function of $\hat{M}$ around a point in $\Sigma_k$ for a fixed $\hat{g}$. For each state one finds a cusp with two branches. In this case all points 
$p$ with $p\in\Sigma_k$ can be understood as centers of the fold catastrophe \cite{saun,dom,pos}. The behavior of $I\left[f_{k,\hat{\omega}} \right]$ as a function of $\hat{\omega}$ close to $ p\in\Sigma_k$ can be described by \[ f_b(x)=x^3 +bx \, ,\] 
where $x$ corresponds to the behavior variable $\hat{\omega}$ and $b$ corresponds to the control parameter $\hat{M}$. For $b=0$, $f_b(x)$ describes the behavior of $I$ at point $p$. For $b<0$, $f_b(x)$ has a maximum and a minimum, which can be 
identified as an unstable and a stable equilibrium solution respectively. For $b>0$, $f_b(x)$ has no real solutions. Regarding that, all points in the upper branches in Fig.~\ref{etotum} correspond to unstable solutions of the GPN system and all 
points in the lower branches correspond to stable solutions of the GPN system. Since points in the lower branches belong to points in the upper part of $\mathcal{M}_k$ in Fig.~\ref{ggm} and vice versa, all points in the upper part of $\mathcal{M}_k$ 
represent stable solutions of the GPN system and all points in the lower part represent unstable solutions respectively.

In Fig.~\ref{ggm} the bifurcation set $\xi_3$ is plotted in the control space $\mathcal{C}=\{\hat{M},\hat{g}\}$. $\xi_3$ and the $\hat{M}$-axis ($\hat{g}=0$) divide $\mathcal{C}$ in three regions. For a fixed pair of control parameters 
$(\hat{M},\hat{g})$ one finds in region S one stable solution, in region SU one stable and one unstable solution and in region N no solution for the third radially excited state of the GPN system.
      
\begin{figure}[floatfix] 
\small
\begin{center}
\input{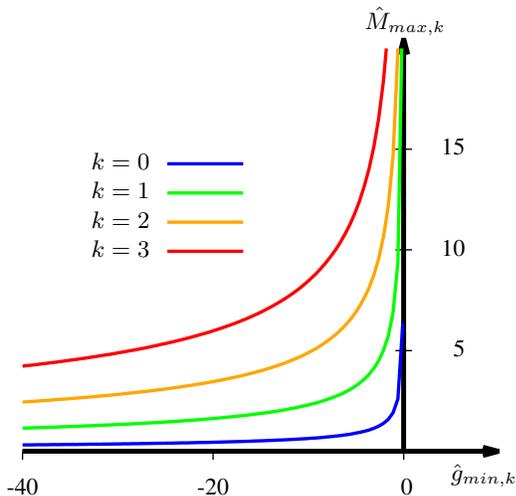}
\caption{Bifurcation set $\xi_k$ of the ground state ($k=0$) and the first three radially excited states ($k=1,2,3$) in the control space $\mathcal{C}$. For all $k$ $\xi_k$ shows the same qualitative shape.}\label{mmaxfa}
\end{center}
\normalsize
\end{figure}  

The bifurcation sets $\xi_k$ of the ground state ($k=0$) and the first three radially excited states ($k=1,2,3$) are plotted in $\mathcal{C}$ in Fig.~\ref{mmaxfa}. $\xi_k$ can be calculated analytically for all states by using the scaling factor 
$\lambda$, as long as for each state the value of one point in $\xi_k$ is known. $\xi_k$ is described by the equation: $\hat{M}_{\text{max},k}(\hat{g})=\sqrt{\hat{g}_k/\hat{g}} \hat{M}_{\text{max},k}$, where $(\hat{M}_{\text{max},k},\hat{g}_k)$ is a 
known point in $\xi_k$. In table \ref{an} the value of $\hat{g}_k$ for $\hat{M}_{\text{max},k}=1$ is given for each state.
 \begin{table}[floatfix]
\begin{center}
\begin{tabular}{ c|c}
	  state                  &$\hat{g}_k$ \\
	  \hline
	  \hline
	  ground state           & -4.1011 \\
	  1. radially excited state & -53.136 \\
	  2. radially excited state & -239.99 \\
	  3. radially excited state & -716.18 \\  
\end{tabular}
\end{center}
\caption{Minimum value of $\hat{g}_k$ in case of $\hat{M}=1$ for the ground state and the first three radially excited states} \label{an}
\end{table}
As shown in Fig.~\ref{mmaxfa}, there are points in $\mathcal{C}$ for which only solutions of higher radially excited states can be found (e.g. in the region between $\xi_3$ of the third radially excited state and $\xi_{2,1,0}$ of the second or first radially excited state or ground state). The radially excited state with the lowest number of nodes has the smallest value of the energy functional, which also leads the smallest total energy $E_{\text{tot}}$.
			
\section{Conclusion} \label{conclusion}

We discussed the ground state and the first three radially excited states of the GPN system with respect to the influence of two external parameters, the total mass $\hat{M}$ and the self-interaction factor $\hat{g}$. The eigenenergy $\hat{\omega}_k$ and the value of the energy functional $I$ are considered as functions of $\hat{M}$ or $\hat{g}$ and they show the same qualitative behavior for the first three radially excited states and the ground state. For large total masses not only the ground state but also the radially excited states approach the TF limit. In the latter case the shape of the envelope of the squared radially excited wave function solutions approaches that of the TF limit.

By using arguments of the catastrophe theory the stability of the ground state and the radially excited states is discussed. For a fixed pair of external parameters one can find either one stable solution ($\hat{g}>0$), one stable and one unstable solution ($\hat{g}<0,\hat{M}<\hat{M}_{\text{max},k}$), or no equilibrium solution at all ($\hat{g}<0,\hat{M}>\hat{M}_{\text{max},k}$) for the ground state and the first three radially excited states. However, considering the results in \cite{cooling2006}, it is highly likely that the radially excited states, which are considered stable in terms of the catastrophe theory, are intrinsically unstable for $\hat{g}>0$ and $\hat{g}<0,\hat{M}<\hat{M}_{\text{max},0}$. This was shown for a few different values of $g$ by F.S. Guzm\'an et al. by studying the time evolution of some radially excited equilibrium solutions, while allowing the flow of particles out of the numerical domain \cite{cooling2006}. 

For $\hat{g}<0$ and $\hat{M}>\hat{M}_{\text{max},0}$ there exists no ground state solution and only solutions of the radially excited states can be found. This leaves the $k$-th radially excited state as the equilibrium solution with the lowest total energy if \begin{equation}
\hat{M}_{\text{max},k}>\hat{M}>\hat{M}_{\text{max},k-1} \, . 
\end{equation} 
In \cite{cooling2006} there is nothing said about the intrinsic stability of the excited states in this case. One might expect stability for the lowest-energy equilibrium solution, since the total mass $\hat{M}$ is conserved. However, as discussed for a general relativistic treatment of self-gravitating BECs \cite{miniboson}, this might not be the case. Since the lowest energy equilibrium state possesses not the lowest energy of all possible configurations a numerical discussion of these solutions in \cite{miniboson} revealed them to be dynamically unstable. A similar discussion regarding dynamical stability for the GPN system remains to be done.


%
%

%

\begin{acknowledgments}
We would like to thank D. Giulini for helpful remarks.
Financial support by the DFG Research Training Group 1620 ``Models of Gravity'' is gratefully acknowledged.
\end{acknowledgments}

\bibliography{sbec_template_cuver4.bib}

\end{document}